\documentstyle[prl,aps,epsf,multicol]{revtex}

\begin{document}

%
\newcommand{\fig}[2]{\epsfxsize=#1\epsfbox{#2}}
%
%
%
 \newcommand{\passage}{
         \end{multicols}\widetext%
                \vspace{-.5cm}\noindent\rule{8.8cm}{.1mm}\rule{.1mm}{.4cm}} 
 \newcommand{\retour}{
         \vspace{-.5cm}\noindent\rule{9.1cm}{0mm}\rule{.1mm}{.4cm}\rule[.4cm]{8.8cm}{.1mm}%
         \begin{multicols}{2} }
 \newcommand{\unecol}{\end{multicols}}
 \newcommand{\deuxcol}{\begin{multicols}{2}}
%

\tolerance 2000

\title{Disordered XY models and Coulomb gases: renormalization
via traveling waves}
\author{David Carpentier and Pierre Le Doussal}
\address{CNRS-Laboratoire de Physique Th{\'e}orique de l'Ecole Normale 
Sup{\'e}rieure, 24 Rue Lhomond, 75231 Paris\cite{lptens}
}

\maketitle

\begin{abstract}
We present a novel RG approach to 2D random 
XY models using direct and
replicated Coulomb gas methods. By including
fusion of environments (charge fusion
in the replicated CG) it follows the distribution of local disorder,
found to obey a Kolmogorov non linear equation (KPP)
with traveling wave solutions. At low T and weak disorder
it yields a glassy XY phase with broad distributions
and precise connections to Derrida's GREM. Finding
marginal operators at the disorder-induced transition
is related to the front velocity selection problem in KPP
equations yielding new critical behaviour.
The method is applied to critical random Dirac problems.
\end{abstract}

\deuxcol


Two dimensional random systems have attracted
considerable recent interest in domains ranging
from localization in quantum Hall systems
to vortices in superconductors.
In the context of localization, progress was made
to characterize the multifractal statistics of 2D wavefunctions
using random Dirac models \cite{chamon96,castillo97},
extending previous studies in 1D \cite{comtet97}.
On the other hand the glassy properties of vortex phases
with disorder was investigated using random
XY models. While these lines of studies
have developed in an apparently disconnected way,
they led to similar proposals \cite{tang96,chamon96} that remarkable
connections exist between the large fluctuation properties
of these systems and Derrida's random energy models GREM
\cite{rem}. To study these connections further, consistent
RG techniques are needed. Our aim is
to develop such an approach, which, as in other 
glassy systems, e.g. in 1D \cite{fisherrg}, requires
a proper treatment of broad distributions.

Here we focus on random gauge XY models
and discuss at the end
related random Dirac problems. 
Recently the phase diagram
predicted long ago in \cite{rubinstein83} was
reexamined\cite{nattermann95,cha95}
using energy arguments. It was proposed that 
defects (vortices)
are not thermalized at low T, leading to a non-reentrant transition
line between the XY and the defective phases,
and to a failure \cite{korshunov93} of the conventional Coulomb Gas perturbative
expansion of \onlinecite{rubinstein83}.
This motivated several interesting proposals for
RG procedures \cite{tang96,scheidl97,korshunov96}.
However, these approaches, while giving the correct
topology of the phase diagram, are not
fully consistent, since they do not take into
account the renormalization of local disorder.
As shown here, this changes quantitatively
the results of \cite{tang96,scheidl97,korshunov96}
for the XY phase and becomes crucial at
the disorder driven transition.

In this Letter we reexamine the RG procedure for
the disordered Coulomb gas (CG) and for the random gauge XY model
and propose an approach which allows to treat the
probability distribution of the local disorder. Technically this
amounts to introduce composite {\it charge} fugacities 
and to study their fusion (in the CG sense) upon coarse graining,
while only {\it dipole} fugacities were considered
previously \cite{tang96,scheidl97,korshunov96}. A precise
connection between the fugacity distribution and 
the GREM free energy distribution is found via the Kolmogorov (KPP) equation
\cite{bramson83}
which arises as its RG flow equation. Universality in
the corresponding non linear front selection problem 
unexpectedly translates into the RG universality
around the disorder driven transition. 
This allows
to describe the phase
dominated - and the transition driven - by rare configurations
of frozen defects, where correlations are broadly distributed.
Restriction to the single charge sector yields
a RG derivation of the multifractal properties of
the critical Dirac wavefunction.

The 2D square lattice XY model with random phases \cite{rubinstein83}
is defined by its partition sum $Z[A]=\prod_{i}\int_{-\pi}^{\pi} 
d \theta_{i}~ e^{-\beta H[\theta,A]}$
with:
\begin{eqnarray}  \label{xy}
\beta H[\theta,A]=\sum_{\langle i,j \rangle} V(\theta_{i}-\theta_{j}-A_{ij}) 
\end{eqnarray}
and $V(\theta)=- \frac{K}{\pi} \cos(\theta)$, $K=\beta J$, $\beta=1/T$.
The $A_{ij}$ are independent gaussian random gauge fields,
with $\overline{A_{ij}^{2}}= \pi \sigma$.
This model can be transformed exactly \cite{footnote:villain,sinegordon}
into a CG with integer charges defined on the sites ${\bf r}$
of the dual lattice with $Z[V] = \sum_{\{n_{\bf r}\}} e^{- \beta H}$
and:
\begin{eqnarray} \label{square}
&& \beta H  = - K \sum_{{\bf r} \neq {\bf r'}} n_{{\bf r}} G_{{\bf r} {\bf r}'}
n_{{\bf r'}} +  \sum_{{\bf r}} n_{{\bf r}} V_{{\bf r}} 
\end{eqnarray}
where $G_{{\bf k}}^{-1}=\frac{1}{\pi} (2-\cos (k_{x}a)-\cos (k_{y}a))$
is the lattice Laplacian.
The bare disorder potential, $V_{{\bf r}} = \frac{K}{\pi} G_{r r'} 
(\nabla \times A)_{\bf r'}$ is gaussian with logarithmic long range
correlations  $\overline{V_{{\bf k}} V_{-{\bf k}}}= 2\sigma K^{2} G_{{\bf k}}$.
The usual continuum approximation with (integer) charges of hard core
$a$ and fugacities $y=e^{-\gamma K}$ of this lattice model
is obtained by using the asymptotic form
$G_{r r'} \approx (\ln |r-r'|/a + \gamma)(1-\delta_{r r'})$.
Here, the
perturbative expansion of $Z[V]$ in $y$, valid in the dilute limit
uniformly over the system, fails.

Before turning to the more systematic replica approach,
let us first sketch the direct RG method suited to
the present case where disorder favors some regions,
resulting in a site dependent local fugacity $y_{\bf r}$.
Our expansion captures the limit where 
the fugacity is negligible almost everywhere
except in a few {\it rare favorable regions}.
This is achieved by following the local disorder 
distribution which is {\it not gaussian}, a novel feature
from all previous approaches. 
We find that the disorder $V_{\bf r}=V^{>}_{\bf r} + v_{\bf r}$
naturally splits into two
parts, a {\it long range correlated gaussian} part $V^{>}_{\bf r}$ with logarithmic 
correlator $\overline{(V^{>}_{\bf r} - V^{>}_{{\bf r}'})^2} 
=4 \sigma  K^2 \ln(|{\bf r}-{\bf r}'|/a)$ and a {\it local non gaussian} part
$v_{\bf r}$ which defines the {\it local fugacity variables}
$z_{\pm}^{\bf r} = y_{\bf r} \exp(\pm v_{\bf r})$ for $\pm 1$ charges
\cite{highercharges}
which have only short range correlations. The 
RG equation for the distribution $P(z_{+},z_{-})$ of
local environments is obtained from two
contributions (i) {\it ``rescaling''}:
upon coarse graining $a \to \tilde{a}=a e^{dl}$, 
$V^{>}$ produces a gaussian additive contribution to
$v$: from $\overline{(V^{>}_{\bf r} - V^{>}_{{\bf r}'})^2}^a
=4 \sigma  K^2 [ \ln(|{\bf r}-{\bf r}'|/\tilde{a}) + dl ]
\equiv \overline{(V^{>}_{\bf r} - V^{>}_{{\bf r}'})^2}^{\tilde{a}}
+ \overline{(dv_{\bf r} - dv_{{\bf r}'})^2} $ one gets
$z_{\pm}^{\bf r} \to z_{\pm}^{\bf r}  
e^{ K dl \pm dv_{\bf r}}$ with $
\overline{dv_{\bf r} dv_{{\bf r}'}}=2 \sigma K^2 dl \delta_{\bf r,\bf r'}$.
(ii) {\it ``fusion of charges''}
(fusion of environments) upon the change of cutoff, two regions with fugacities
$z^{{\bf r}'}_{\pm},z^{{\bf r}''}_{\pm}$ are replaced by a single
region at $\tilde{\bf r}=\frac{1}{2}({\bf r}' + {\bf r}'')$
of effective fugacities $\tilde{z}_{\pm}=(z^{{\bf r}'}_{\pm} + z^{{\bf r}''}_{\pm})/(1 +
z^{{\bf r}'}_{+} z^{{\bf r}''}_{-} + z^{{\bf r}'}_{-} z^{{\bf r}''}_{+} )$
obtained from the relative weight
$W_{+}/W_{0}$ of a charge 1 configuration
(either in ${\bf r}'$ or ${\bf r}''$)
versus a neutral one (either no charge or a dipole). (i) and (ii)
yield:
\passage 
\begin{eqnarray}  \label{rgeqp}
&& \partial_l P(z_+,z_-) =  {\cal O} P - 2 P(z_+,z_-)  + 
 2 \left<  \delta( z_+ - \frac{z'_+ + z''_+}{1 + z'_-  z''_+ + z'_+  z''_-})
\delta( z_- - \frac{z'_- + z''_-}{1 + z'_-  z''_+ + z'_+  z''_-}) 
 \right>_{P'P''}  
\end{eqnarray}
\retour 
where $\langle A \rangle_{P' P''}$ denotes $\int_{z'_{\pm},z''_{\pm}}
A~P(z'_+,z'_-) P(z''_+,z''_-)$ and
${\cal O}= K (2 + z_+ \partial_{z_+} +
z_- \partial_{z_-}) 
+ \sigma K^2 (z_+ \partial_{z_+} - z_- \partial_{z_-})^2$
is the diffusion operator \cite{rguniversal}.

To put this derivation on a firmer footing and to capture broad distributions
of local fugacities we
introduce \cite{inprep} an {\it expansion of physical quantities
in the number of points} \cite{expansion}, which for the free energy
$F[V]=-T \ln Z[V]$ as a {\it functional} of the disorder
reads:
\begin{eqnarray}
\label{free}
F[V]&=& \sum_{{\bf r}_{1}\neq {\bf r}_{2}}
f^{(2)}_{{\bf r}_{1},{\bf r}_{2}}[V]  
+ \sum_{{\bf r}_{1}\neq {\bf r}_{2}\neq {\bf r}_{3}} 
f^{(3)}_{{\bf r}_{1},{\bf r}_{2},{\bf r}_{3}}[V]
+\dots
\end{eqnarray}
where by definition $f^{(k)}_{{\bf r}_{1},..{\bf r}_k}$ depends only 
\cite{taylor} on
$V({\bf r})$ at points ${\bf r}_i$, $i=1,..k$. The continuum
limit of (\ref{free}) consists in replacing the sums
by integrals with hard core constraints
around each ${\bf r}_i$. The first two terms read:
\begin{eqnarray}  \label{twoterms}
&& - \beta f^{(2)}_{{\bf r},{\bf r}'} = \ln(1 + W_{{\bf r},{\bf r}'}) , \qquad
W_{{\bf r},{\bf r}'}=w_{{\bf r},{\bf r}'} + w_{{\bf r}',{\bf r}} \\
\label{thirdorder}
&& - \beta f^{(3)}_{{\bf r}_{1},{\bf r}_{2},{\bf r}_{3}} = \ln( \frac{1 + W_{{\bf r}',{\bf r}''} 
+ W_{{\bf r},{\bf r}'} + W_{{\bf r},{\bf r}''}}{
(1 + W_{{\bf r},{\bf r}'})(1 + W_{{\bf r}',{\bf r}''})
(1 + W_{{\bf r},{\bf r}''})})
\end{eqnarray}
where $w_{{\bf r},{\bf r}'}= e^{-V_{\bf r}+V_{{\bf r}'}-G_{{\bf r},{\bf r}'}}= 
z_{+}^{{\bf r}} z_{-}^{{\bf r}'} e^{-V^{>}_{{\bf r}}+V^{>}_{{\bf r}'}
-G_{{\bf r},{\bf r}'}}$ is the 
Boltzmann weight of a dipole. The first term (\ref{twoterms}) 
corresponds to the independent dipole approximation 
\cite{korshunov96} while (\ref{thirdorder}) takes into account
contributions from triplets of sites. Though
there are {\it no actual configuration with
three charges} in a given environment (neutrality),
this term (and higher orders), overlooked in previous 
approaches, is crucial for
the renormalization as it leads to fusion of
environments when coarse graining. Upon increase
of the cutoff $a \to a e^{dl}$
on the continuum version of (\ref{free}),
the $k$ point integral gives a correction 
to order $dl$ to the $k-1$ point integral, e.g.
$\int_{a<|{\bf r}'-{\bf r}''|<a e^{dl}} f^{3}_{{\bf r},{\bf r}',{\bf r}''}$
corrects $f^{2}_{{\bf r},\tilde{{\bf r}}}$ 
(with $\tilde{{\bf r}}=\frac{1}{2}({\bf r}'+{\bf r}'')$)
as $\delta \ln(1 + W_{{\bf r},\tilde{{\bf r}}}) = dl 
[ \ln(1 + \tilde{W}_{{\bf r},\tilde{{\bf r}}}) -
\ln(1 + W'_{{\bf r},\tilde{{\bf r}}})  - \ln(1 + W''_{{\bf r},\tilde{{\bf r}}}) ]$.
First $W_{{\bf r}' {\bf r}''} = z'_{+} z''_{-} + z'_{+} z''_{-}$
since $V^{>}_{{\bf r}'} = V^{>}_{{\bf r}''} + O(dl)$
(we denote $z_{\pm}^{{\bf r}'}=z'_{\pm}$
etc..), thus the
combination $(w_{{\bf r},{\bf r}'} + w_{{\bf r},{\bf r}''})/
(1 + W_{{\bf r}'{\bf r}''})$ 
in (\ref{thirdorder}) 
can be rewritten as $\tilde{w}_{{\bf r},\tilde{{\bf r}}}
= z_{+} \tilde{z}_{-} e^{-V^>_{\bf r} + V^>_{\tilde{{\bf r}}} 
- G_{{\bf r},\tilde{{\bf r}}}}$
in terms of the new fugacity $\tilde{z}_{\pm}$
defined above. Second, $w'_{{\bf r},\tilde{{\bf r}}}
= z_{+} z'_{-} e^{-V^>_{\bf r} + V^>_{\tilde{{\bf r}}}
-G_{{\bf r},\tilde{{\bf r}}}}$ 
(similarly $w'_{{\bf r},\tilde{{\bf r}}}$ using $z''_{\pm}$).
Averaging this correction over disorder yields
a result corresponding to the RG equation
for $P$ (\ref{rgeqp}). A complete derivation
of (\ref{rgeqp}) involves a similar procedure
for all moments of $F[V]$ \cite{inprep},
provided one adds to (\ref{free}) the free energy sum of all
degrees of freedom eliminated
up to scale $l$ with
$\partial_l F_0 =- T \langle \ln(1+z'_{+} z''_{-} + z'_{-} z''_{+}))
\rangle_{P_l'P_l''}$. 

Finally, the 
renormalization of $K$ and $\sigma$ is obtained from screening
\cite{tang96,scheidl97,korshunov96}:
$\partial_{l}K^{-1}=
-2\pi^{2}\overline{\langle n_{0}n_{R=a}\rangle_{c} }, 
\partial_{l} \sigma =
-2\pi^{2} \overline{\langle n_{0}\rangle \langle
n_{R=a}\rangle}$. Expanding in the number of
sites using (\ref{free}), we obtain \cite{inprep} the following,
which together with (\ref{rgeqp}) forms our complete
set of RG equations \cite{nofusion,cd}
\begin{mathletters} \label{screening}
\begin{eqnarray} 
&& \frac{dK^{-1}}{dl} = \frac{4 d'}{d^2}
\left< \frac{ z'_+ z''_- + z'_- z''_+ + 4 z'_+ z''_- z'_- z''_+ }
{(1 + z'_+ z''_- + z'_- z''_+)^2} \right>_{P' P''} \\
&& \frac{d\sigma}{dl} = \frac{4 d'}{d^2} 
\left< \frac{ (z'_+ z''_-  - z'_- z''_+)^2 }{
(1 + z'_+ z''_- + z'_- z''_+)^2}  \right>_{P' P''}
\end{eqnarray}
\end{mathletters}

The combinatorics necessary to this method is
much easier performed using replicas. 
We start again from (\ref{square})
and represent $Z^m$ as the partition sum of 
a CG with m-{\it vector} charges $n^b_{\bf r}$ living
on the dual lattice sites.
Averaging over disorder, and taking the continuum 
approximation we obtain the m-vector (hard core) CG of 
partition sum expanded in power of the {\it vector fugacity}
$Y_{\bf n}$:
\begin{eqnarray*} 
&& \overline{Z^m} = 1 + \sum_{p \geq 2}
\sum_{{\bf n}_1..{\bf n}_{p}} 
\int_{{\bf r}_1..{\bf r}_p}
Y_{{\bf n}_1}..Y_{{\bf n}_p}
\prod_{i \neq j} \left|\frac{r_i-r_j}{a}\right|^{ n^b_i K_{b c} n^c_j }
\end{eqnarray*}
with $K_{bc}=K \delta_{bc} - \sigma K^2$, all
integrals being restricted to $|r_i-r_j|>a$, and the sum is
over all distinct neutral configurations $\sum_{\bf r} n^b_{\bf r}=0$.
$Y_{\bf n}$ is a function of the
replicated charge ${\bf n}=(n^1,..n^m)$ with 
bare value $Y_{\bf n} \approx e^{-\gamma n_b K^{bc} n_c}$.
Since $K_{b \neq c} \neq 0$, one cannot restrict to 
single non zero component charges \cite{scheidl97},
as it leads to the erroneous results of \cite{rubinstein83}
at low temperature. However, we stress that
this quadratic form for $Y_{\bf n}$, which results from
the Gaussian nature of the bare disorder, is {\it not} preserved by the RG
as shown below. We now perform the RG analysis of the m-vector
CG, extending the scalar case \cite{nienhuis87},
leaving the above form unchanged 
with \cite{cd}:
\begin{eqnarray}   \label{rgrep}
&& \partial_l K^{-1}_{bc} = d' \sum_{n \neq 0} n^b n^c Y_{\bf n}
 Y_{-{\bf n}} \\ 
&& \partial_l Y_{\bf n \neq 0} = ( 2-  n^b K_{bc} n^c ) Y_{\bf n}
+ d  \sum_{{\bf n}' \neq 0, {\bf n}} 
Y_{{\bf n}-{\bf n}'} Y_{\bf n'}  \label{rgrep2}
\end{eqnarray}
Rescaling and 
annihilation of opposite replica charges
separated by 
$a \leq |{\bf r}_{i}-{\bf r}_{j}| \leq ae^{dl}$
gives the first term of (\ref{rgrep2}) and (\ref{rgrep}).
The second term of (\ref{rgrep2}) which comes from {\it fusion of
two replica charges} as usual in {\it vector}
CG, was absent in \cite{tang96,korshunov96,scheidl97}
but is necessary for consistency of RG to
order $Y_{\bf n}^2$.

Why should one consider the expansion in $Y_{\bf n}$ ?
Technically, it is valid, together with 
(\ref{rgrep},\ref{rgrep2}), in the limit of a small density of 
{\it vector charges} \cite{nienhuis87}, which
here corresponds to a small density of favorable
local regions. Indeed we checked
that this expansion is {\it identical} term by term, for
$m \to 0$, to the expansion in number of points 
of the free energy (\ref{free}). 
Thus the set of $Y_{\bf n}$ should encode the
full scale dependent distribution $P(z_{\pm})$ of local disorder,
the perturbative parameter being $P(z_{+} \sim 1)$.
Remarkably, the correspondence between  $P(z_{+},z_{-})$ and $Y_{\bf n}$
emerges when performing the analytical continuation $m \to 0$
of (\ref{rgrep},\ref{rgrep2}) which we now present. To capture the
most relevant operators it is sufficient to consider 
$Y_{\bf n}$ with $n^b = 0, \pm 1$ in each replica \cite{highercharges},
which, using replica permutation symmetry,
depends only on the numbers $n_{\pm}$ of $\pm 1$
components of $n$. This leads to 
the general parametrisation in term of
a function $\Phi(z_{+},z_{-})$:
\begin{eqnarray*}
Y_{\bf n} = \langle z_{+}^{n_{+}}z_{-}^{n_{-}} \rangle_\Phi = \langle
\prod_{b}\left[\delta_{n^{b},0}+z_{+}\delta_{n^{b},+1}+z_{-}\delta_{n^{b},-1}
\right] \rangle_\Phi 
\end{eqnarray*}
where $\langle .. \rangle_\Phi = \int_{z_{\pm}} .. \Phi (z_{+},z_{-})$.
After some combinatorics \cite{cutoff}
the limit $m \to 0$ of
(\ref{rgrep2}) can be rewritten {\it equivalently}
as an equation for $\Phi$, detailed in \cite{inprep}. 
The first term in (\ref{rgrep2}) gives a diffusion
contribution $(2 + {\cal O})\Phi$ and the second term in (\ref{rgrep2})
yields a term of fusion of environments, analogous to the
one in (\ref{rgeqp}). From this equation 
${\cal N}=\int_{z_{-},z_{+}>0} \Phi (z_{\pm})$ is found to satisfy
$\partial_l {\cal N} = 2 {\cal N} - d {\cal N}^2$, and thus
converges quickly \cite{nofusion} towards ${\cal N}^*=2/d$.
We thus define
the normalized $P= \Phi/{\cal N}$
which satisfies (\ref{rgeqp}) and is naturally interpreted as
a probability distribution \cite{rguniversal}. Finally,
with the same definitions, (\ref{rgrep}) yields
(\ref{screening}).

\begin{figure}[thb]  \label{fig1}
\centerline{\fig{8cm}{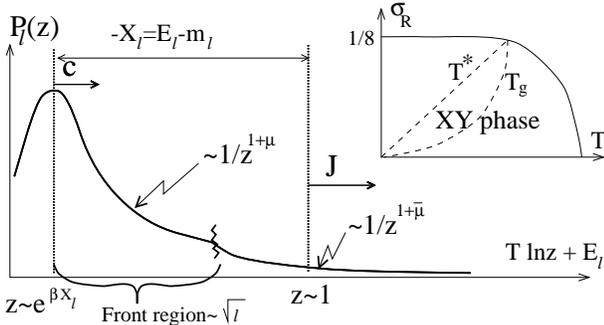}}
\caption{\narrowtext Scale dependent distribution $P_{l} (z)$
and its two tail regions for $T<T_{g}$. Inset: phase diagram}
\end{figure}

We first study numerically the RG equations (\ref{screening},\ref{rgeqp})
and find at low T, $\sigma<\sigma_c$ an XY phase as in Fig.1
($K$, $\sigma$ converge to $K_R$, $\sigma_R$). 
The typical $z$ goes to zero but $P$ develops a broad tail up to
$z \sim O(1)$. While in this phase and at criticality the concentration of
rare favorable regions $P_l(1)$ decreases, it eventually increases
at large $l$ in the disordered phase $\sigma_c  \lesssim \sigma$.

To go beyond numerics, we argue \cite{footnote:P(0)}
that it is consistent to discard the $z'_{+} z''_{-}$ terms in 
the denominators in (\ref{rgeqp}). This leads to
a closed equation for a single
fugacity distribution $P(z)=\int_{z_{-}} P(z,z_{-})$ 
which, using the parametrization 
$G_l(x)=1 - \left< \exp( - z  e^{- \beta (x - E_l)}  ) \right>_{P_l(z)}$
where $E_l=\int_0^l J_l dl$, $\beta=1/T$,
can be rewritten as:
\begin{eqnarray}  \label{kpp}
\frac{1}{2} \partial_l G = D_l \partial_x^2 G + (1-G) G
\end{eqnarray}
The diffusion coefficient is 
$D_l=\frac{1}{2} \sigma_l J_l^2$ and by construction
$G_l(-\infty)=1$ and $G_l(+\infty)=0$.
Remarkably, for constant $D$ this is
the much studied KPP equation, which describes diffusive
invasion of an unstable
state ($G=0$) by a stable one ($G=1$), also related
to branching diffusions and glassy REM-like models
\cite{cayley}. It is known \cite{bramson83,kpp} that $G_l(x)$
converges at large $l$ 
towards traveling waves solutions $h(x-m_{l})$
selected by the behaviour at infinity of $G_{l=0}(x) \sim e^{-\beta x}$.
This implies $P_{l} (z) \to z^{-1}p(\ln z - \beta X_{l})$ 
with $X_{l}=m_{l}-E_{l}$ ($X_l<0$ in the XY phase,
see Fig.1).

At low T in the XY phase, $P_l(z)$ becomes very broad
and one must distinguish
two different tails. As shown in Fig.1, 
the  bulk of the distribution (typical values) is located
around $z_{typ} \sim e^{\beta X_l}$. It corresponds to the 
{\it front region} which has a tail of size
$\sqrt{l}$ ahead of the front. There 
from the velocity selection studies\cite{bramson83,kpp}
for $T>T_g=J \sqrt{\sigma /2}$ we find the front position
$m_{l}\sim 2 (\beta^{-1}+D\beta)l$. 
For $T<T_g$ the velocity freezes with 
$m_{l}=\sqrt{D} (4l -\frac{3}{2}\ln l +O (1))$
and $h_{l} (y)\sim \frac{y}{\sqrt{D}}\exp ( -\frac{y}{\sqrt{D}}-\frac{y^{2}}{8Dl})$.
This corresponds to $P_l(z) \sim z^{(-1+\mu)}$ within the
tail of the front, with $\mu=T/T_g<1$.
Thus for $T<T_g$ the distribution function of $\ln z$ travels at the 
{\it relative velocity} $\partial_{l} X_l= J(\sqrt{8\sigma}-1)$, which
determines the phase diagram: it is
negative (decrease of P(1)) in the low T XY regime, positive for
$\sigma\geq \sigma_{c}$. Furthermore, at low T,
there is also a {\it far tail} ahead $\sim l$ of the front which
corresponds to rare events $z\sim 1$, of small probability $P_l(1)$,
but which dominate average correlations (and thus 
$\partial_l K$ and $\partial_l \sigma$). The linearized KPP equation,
valid in this region, leads to $P_l(z) \sim P_l(1) z^{-1+\overline{\mu}}$
with $\overline{\mu}=T/T^* <1$ and to $P_l(1) \sim
e^{(2-\frac{1}{4 \sigma})l}$, for $T<T^*=2 \sigma J$.
A more detailed study of the XY phase
is given in \cite{inprep}. 

In the high T regime of the XY phase, $P_l(z)$ is not so broad,
and one recovers from (\ref{kpp}) the usual
RG result \cite{rubinstein83} $\partial_{l} y = (2 - K + \sigma K^2) y$
for the average fugacity $y_l= \langle z \rangle_l < +\infty$ ($\sim z_{typ}$ for
$T>T_g$), using $G_l(x) \sim e^{-\beta (x-E_l)} \langle z \rangle_{P_l}$
at large $x$.

The critical behaviour at the transition from the XY to the
disordered phases is determined by {\it the front region},
since the velocity $\partial_{l} X_l$ vanishes.
Here we sketch the analysis at $T=0$: defining $z=e^{\beta u}$
and using $\lim_{\beta \to \infty} \beta p_l(\beta(u-X_l)) = - h'_l(u-X_l)$,
(\ref{screening}) yields:
\begin{eqnarray}
&& \partial_l J^{-1} = - \frac{8 d'}{d^{2}} h'_l(-2 X_{l}) ~~;~~
\partial_l \sigma = \frac{8 d'}{d^{2}} h(-2 X_l)
\end{eqnarray}
(which is of order $P(1)^2$ perturbative for 
$X_{l} \gg 1$). From the {\it universal} corrections \cite{bramson83,kpp}
to the velocity: $\partial_l X_{l}=J (\sqrt{8\sigma}-1)-3\sqrt{D}/2l +O
(l^{-\frac{3}{2}})$ we obtain a projection of the RG flow 
on the plane $\sigma \sim \sigma_{c}=\frac{1}{8}$ and $g \sim P_l(1) \sim h (-X_{l})$ 
which reads \cite{diffusion}:
\begin{eqnarray}  \label{rg2p}
&& \partial_l g = (16(\sigma-\sigma_{c}) - \frac{3}{2 l}) g ~~;~~
 \partial_l (\sigma-\sigma_{c}) = g^2
\end{eqnarray}
yielding $g_{l}\sim l^{-\frac{3}{2}}$ at criticality
and a correlation length 
$\xi\sim e^{\frac{1}{|\sigma-\sigma_{c}|}} $. This new
universality class \cite{inprep} is different from KT and from
the prediction of \cite{scheidl97,tang96}.
Note that although most details of $P(z_{\pm})$ e.g. its bulk,
depend on the cutoff procedure (and fusion rule...), here
the universality appears in a remarkable way. It comes
from the independence of the velocity and the front tail
(which also determine the relevant operators)
on the precise form $F[G]$ of the non linear
term in (\ref{kpp}) (see \cite{bramson83}).

Finally, our RG also applies to the problem of a {\it single charge} $Z=\sum_r e^{-V_r}$
in a random potential  with logarithmic correlations, related to diffusion in random media
\cite{comtet97} or wavefunction of 2D Dirac fermions in a random
magnetic field \cite{chamon96,castillo97,comtet97}. The same decomposition
of disorder, and fusion of environments ($z=z'+z''$, fugacities being
local partition sums) yields (\ref{kpp}). We recover for $P_l(z)$ the
mapping to directed polymers (DP) on Cayley trees, conjectured in \cite{chamon96,tang96},
with the same universal intensive free energy and
wavefunction multifractal spectrum \cite{castillo97,inprep}.

To conclude, we developed a RG
approach to random XY models, disordered CG and random Dirac
problems. By following the whole fugacity distribution, it appears 
perturbative in the concentration of rare favorable
regions, which corresponds to 
the vector fugacity in the replica method. This expansion is highly non perturbative
in the original fugacity $y$.
A precise connection to the free energy distribution
of DP on Cayley trees
and GREM arises from the RG \cite{rsb} and turns out to be crucial
to describe the disorder driven transition.

We thank B. Derrida and V. Hakim for useful discussions
about the KPP equation.

\unecol

\begin{thebibliography}{99}

\bibitem[**]{lptens}
Laboratoire propre associ{\'e} {\`a} l'ENS et {\`a} Univ. Paris-Sud.

\bibitem{chamon96} C. de C. Chamon, C. Mudry and X. G. Wen.
Phys. Rev. Lett. {\bf 77} 4194 (1996).

\bibitem{castillo97} H. Castillo et al. Phys. Rev. B. {\bf 56}
10668 (1997).

\bibitem{comtet97} For review see A. Comtet, C. Texier
cond-mat/9707313, Lect. notes in Phys. 502 (1998).

\bibitem{tang96} L. H. Tang, Phys. Rev. {\bf B 54}, 3350 (1996).

\bibitem{rem} B. Derrida Phys. Rev. B {\bf 24} 2613 (1981),
J. Phys. Lett. {\bf 46} 401 (1985).

\bibitem{fisherrg} D. S. Fisher, Phys. Rev. B {\bf 50} (1994),
D. S. Fisher, P. Le Doussal, C. Monthus cond-mat/9710270.

\bibitem{rubinstein83} M. Rubinstein et al., Phys. Rev. {\bf B 27}, 1800 (1983)

\bibitem{nattermann95} T. Nattermann et al., J. Phys. I (France) {\bf 5}, 
565 (1995)

\bibitem{cha95}M. Cha and H.A. Fertig, Phys. Rev. Lett. {\bf 74}, 4867 (1995)

\bibitem{korshunov93} S. E. Korshunov, Phys. Rev. {\bf B 48}, 1124 (1993)

\bibitem{scheidl97} S. Scheidl, Phys. Rev. {\bf B 55}, 457 (1997)

\bibitem{korshunov96} S.E. Korshunov and T. Nattermann, Phys. Rev. {\bf B 53},
2746 (1996)

\bibitem{bramson83}
M. Bramson Mem. Am. Math Soc. {\bf 44} No 285 (1983).

\bibitem{footnote:villain} it is exact for the
Villain form $e^{-V(\theta)}=\sum_p
e^{-\frac{K}{2 \pi} (\theta-2 \pi p)^2}$
which should be in the same universality
class.

\bibitem{sinegordon}
As will appear here (\ref{square}) is
equivalent to the random Sine Gordon model
$H=\int_r (\nabla \phi)^2 + i {\bf a} \cdot \nabla \phi + z_{+} e^{i \phi} +
z_{-} e^{- i \phi}$ with a natural splitting of disorder ${\bf a}, z_{\pm}$.
The replica OPE of $Y_{\bf n} e^{i {\bf n} \cdot {\bf \Phi} }$ also yields
(\ref{rgrep},\ref{rgrep2}).

\bibitem{highercharges}
higher charges, e.g. $\pm 2$, are less relevant since
the diffusion operator for $\int_{z_{\pm}} P(z_{\pm},z_{++},z_{--})$
is as in (\ref{rgeqp}) with $K \to 4 K$ 
and $\sigma \to 2 \sigma$ and fusion leads to
$P(z_{++} \sim 1) \sim P(z_{+} \sim 1)^2$.

\bibitem{inprep}
D. Carpentier, P. Le Doussal in preparation.

\bibitem{expansion} i.e. in the number of
favorable regions, $P_l(1)$.

\bibitem{taylor}
a Taylor expansion shows \cite{inprep} that
$f^{(k)}({\bf r}_{1} \dots {\bf r}_{k}) =
 \sum_{l=0}^{l=k} (-1)^{k-l} \sum_{i_{1},\dots,i_{l}\in [1\dots k]}
 F^{l}_{{\bf r}_{i_{1}},\dots,{\bf r}_{i_{l}}}$
in terms of the $l$ site free energy $F^{l}_{{\bf r}_{i_{1}},\dots,
{\bf r}_{i_{l}}}$. 

\bibitem{nienhuis87} B. Nienhuis, in {\it Phase transitions and critical
phenomena}, Domb Green Ed, Vol. 11 (1987)

\bibitem{cutoff}
(\ref{rgeqp})  corresponds to a given branching
process, associated to a particular cutoff,
which even at $\sigma=0$ contains disorder 
in the positions of the branching nodes. The most appropriate
cutoff would yield $Y[{\bf n}] =y^{{\bf n}^2}$ corresponding to
a non linear term $F[G]=(G-1) \ln(1-G)$ in (\ref{kpp}). 

\bibitem{nofusion}
Without fusion our diffusion
formalism yields $z_{\pm}=y_l e^{\pm v_l}$ where 
$v_l$ is gaussian. Redefining dipole fugacity
via $v=v'+v''$ we recover 
(26) of \cite{scheidl97}(using ${\cal N} \sim e^{2 l}$).

\bibitem{cd} $d'=2 \pi^2$, $d=\pi$ for our cutoff choice
with $d'/d^2$ universal. The cubic term in $\partial_{l}Y $ in
\cite{nienhuis87} drops out for $m\rightarrow 0$. 

\bibitem{rguniversal}
The fraction
of fusioned environments yields the universal factor
$\partial_{l} V/V =2$.

\bibitem{footnote:P(0)}
This amounts to neglect terms of order $P(1)^3$ in
the RG equation for $P(1)$
and in (\ref{screening}) since $P(z_{+} \sim 1,
z_{1} \sim 1)$ consistently remains of order $P(1)^2$.

\bibitem{cayley} B. Derrida, H. Spohn 
J. Stat. Phys. {\bf 51} 817 (1988).

\bibitem{kpp} E. Brunet and B. Derrida Phys. Rev. E {\bf 56} 2597
(1997). U. Ebert and W. Van Sarloos, Phys. Rev. Lett. Feb. 1998.

\bibitem{diffusion}
In (\ref{rg2p}) subdominant contributions are neglected, e.g
the variations of $D_l$.

\bibitem{rsb} Direct replica solution \cite{rem} of GREM models 
requires replica symmetry breaking (RSB) for $T<T_g$. It
yields exponential free energy distributions (Boltzman weights
$z_{\bf r}/\sum_{{\bf r}'} z_{{\bf r}'}$ being
dominated by a few states, a characteristic of RSB).
Within the RG it translates into generation of
broad distributions $P(z) \sim z^{-(1+\mu)}$
($\mu<1$, no first moment) which have similar
properties. This suggests to follow broad distributions alternatively
in a more conventional RG (with gaussian distributions)
but with RSB \cite{inprep}.


\end{thebibliography}
\end{document}